# Machine Learning of Mechanical Properties of Steels

Jie XIONG[b,c], Tong-Yi ZHANG[a,*], San-Qiang SHI[b,c,*]

[a]Materials Genome Institute, Shanghai University, Shanghai, China

[b]Department of Mechanical Engineering, The Hong Kong Polytechnic University, Hong Kong, China

[c]Shenzhen Research Institute, The Hong Kong Polytechnic University, Shenzhen, China

* corresponding author: T.-Y. ZHANG (zhangty@shu.edu.cn); S.-Q. SHI (san.qiang.shi@polyu.edu.hk).

**Abstract:** The mechanical properties are essential for structural materials. The analyzed 360 data on four mechanical properties of steels, viz. fatigue strength, tensile strength, fracture strength, and hardness, are selected from the NIMS database, including carbon steels, and low-alloy steels. Five machine learning algorithms were applied on the 360 data to predict the mechanical properties and random forest regression illustrates the best performance. The feature selection was conducted by random forest and symbolic regressions, leading to the four most important features of tempering temperature, and alloying elements of carbon, chromium, and molybdenum to the mechanical properties of steels. Besides, mathematic expressions were generated via symbolic regression, and the expressions explicitly predict how each of the four mechanical properties varies quantitatively with the four most important features. The present work demonstrates the great potential of symbolic regression in the discovery of novel advanced materials.

## 1. Introduction

Seeking structure-property relationships is the scientifically fundamental and best approach to new materials discovery. Comprehensively understanding and controlling the structure-property relationships are a great challenge due to the diversity and complexity of materials. Data-driven discovery of novel advanced materials utilizes the emerging science and technology from big data and artificial intelligence (AI), data mining, and machine learning to accelerate materials research and development [1–8]. Materials data and machine learning provide the foundation of the data-driven materials discovery paradigm. The approach integrated materials domain knowledge and AI technology forms a new research field, materials informatics. The Materials Genome Initiative (MGI) aims at the half cost and half period time to accomplish the entire course from discovery, development, to deployment of advanced materials [9]. With the integrated approach, materials data are used to explore structure-property relationships and to build up models and guidance for new materials synthesis. For example, Homer et al. [10] and Zhu et al. [11] employed Machine Learning (ML) tools to investigate grain boundaries in polycrystalline materials. Raccuglia et al. [12] demonstrated a ML strategy to elucidate how to classify the successful and failed synthesis conditions by using historically accumulated experimental data. Agrawal et al. [13,14] used ML algorithms to predict the fatigue strength of steels, which made a substantial impact in the understanding of fatigue behavior. However, their ML predictions did not result in explicit mathematic expressions between features and output properties which are much more desirable in materials research, design, development, and deployment.

The purpose of this work is to predict the four mechanical properties of steels via five ML algorithms, especially, the algorithms of Random Forest (RF) regression and Symbolic Regression (SR). The performances of the five algorithms are assessed and RF is the best, and the explicit mathematic expressions are obtained from SR.

**2. Data resource**

The present work uses the publicly available dataset for steels in Japan National Institute of Material Science (NIMS) [15], which is one of the most massive experimental datasets in the world. The NIMS datasets contain materials chemical compositions, processing conditions, and property information including the mechanical properties of steels at room temperature, such as fatigue strength, tensile strength, fracture strength, and hardness. Fatigue strength is defined as the critical value of applied stress range, at and below which no fatigue failure will occur at a given fatigue life. The rotating bending fatigue strength, called fatigue strength hereafter for simplicity, at fatigue life of $10^7$ cycles was used in this work.

Fatigue testing conditions of loading frequency and profile, testing temperature and environment, and specimen dimensions etc. have significant effects on fatigue behavior. The 393 original data collected from NIMS database were all fatigued with same testing conditions so that the testing conditions will not be considered in the present work. The 393 original fatigue samples are consisted of 113 carbon steels, 258 low-alloy steels, and 22 stainless steels, described with chemical compositions, processing parameters, inclusion parameters, and mechanical properties. The compositions include nine alloying elements (C, Si, Mn, P, S, Ni, Cr, Cu, Mo). The inclusion parameters are the area fraction of non-metallic inclusions, including dA (inclusions formed during plastic work), dB (inclusions occurring in discontinuous arrays), and dC (isolated inclusions). The processing parameters are the reduction ratio from the ingot to the bar, and heat treatment parameters described in detail below.

(1) The heating rate and cooling rate are not considered here because of no such kind of data.

(2) Three types of heat treatments, viz. normalizing, quenching, and tempering, were conducted on the steels. The temperatures of normalizing, quenching, and tempering are included in the data, while the holding times at heat treatment temperatures are not considered in the present work, because only two holding times are available.

(3) After the heat treatment, the samples were cooled down to room temperature and the fatigue tests were conducted at room temperature.

(4) 11 carbon steels without normalizing treatment (SC25 steels), and 22 stainless steels without quenching and tempering treatment are excluded from the present study, which reduces the original 393 data to 360.

Table 1. 16 Features of the 360 NIMS fatigue data

| Features | Description | Min | Max | Mean | StdDev |
|---|---|---|---|---|---|
| NT | Normalizing Temperature (°C) | 825 | 900 | 865.6 | 17.37 |
| QT | Quenching Temperature (°C) | 825 | 865 | 848.2 | 9.86 |
| TT | Tempering Temperature (°C) | 550 | 680 | 605 | 42.4 |
| C ($x_1$) | wt% of Carbon | 0.28 | 0.57 | 0.407 | 0.061 |
| Si ($x_2$) | wt% of Silicon | 0.16 | 0.35 | 0.258 | 0.034 |
| Mn ($x_3$) | wt% of Manganese | 0.37 | 1.3 | 0.849 | 0.294 |
| P ($x_4$) | wt% of Phosphorus | 0.007 | 0.031 | 0.016 | 0.005 |
| S ($x_5$) | wt% of Sulphur | 0.003 | 0.03 | 0.014 | 0.006 |
| Ni ($x_6$) | wt% of Nickel | 0.01 | 2.78 | 0.548 | 0.899 |
| Cr ($x_7$) | wt% of Chromium | 0.01 | 1.12 | 0.556 | 0.419 |
| Cu ($x_8$) | wt% of Copper | 0.01 | 0.22 | 0.064 | 0.045 |
| Mo ($x_9$) | wt% of Molybdenum | 0 | 0.24 | 0.066 | 0.089 |
| RR | Reduction ration | 420 | 5530 | 971.2 | 601.4 |
| dA | Plastic work-inclusions | 0 | 0.13 | 0.047 | 0.032 |
| dB | discontinuous array-inclusions | 0 | 0.05 | 0.003 | 0.009 |
| dC | isolated inclusions | 0 | 0.04 | 0.008 | 0.01 |

*The weight percentage of iron is $x_{10} = 100 - \sum_{i=1}^{9} x_i$

Then, the original 393 data are reduced to 360 data and each datum includes 16 variables of nine alloying elements, one reduction ratio, three heat treatment temperatures, three inclusions, and four target properties (fatigue strength, tensile strength, fracture strength, and hardness). The 16 variables are named features in ML and the minimum and maximum values of each feature are shown in Table 1.

## 3. Results and discussion

### 3.1 ML Models with All Features

Four ML algorithms including RF, linear least square (LLS), k-nearest neighbours (KNN), and architecture-neutral network (ANN) were conducted on the dataset with all 16 features (termed as All). Their performances were evaluated by ten-folds cross-validation. In the ten-folds cross-validation, the data was divided into ten parts, nine parts as training data and one part as testing data, and cycling the training and testing ten times to let all data be used in testing. The predictive power of a ML model on the testing data are measured by the correlation coefficient (R), and the relative root mean square errors (RRMSE), which are defined by

$$R = \frac{|\sum_{i=1}^{n}(y_i-\bar{y})(\hat{y}_i-\bar{\hat{y}}_i)|}{\sqrt{\sum_{i=1}^{n}(y_i-\bar{y}_i)^2 \sum_{i=1}^{n}(\hat{y}_i-\bar{\hat{y}}_i)^2}}, \qquad (1)$$

$$RRMSE = \frac{\sqrt{\frac{1}{n}\sum_{i=1}^{n}(y_i-\hat{y}_i)^2}}{\bar{y}_i}, \qquad (2)$$

where $n$ is the number of testing data, $y$, $\hat{y}$, and $\bar{y}$ denotes the actual value, predicted value, and average value, repressively. R is between 0 and 1, and a value of 1 indicates a perfect prediction. The RRMSE value of zero indicates a perfect fit. In general, higher R and lower RRMSE indicates a better ML model [16].

Figure 1 shows the R and RRMSE values of the four ML models and illustrates the best predicted values, from one of the ML models, against the measured values for each of the four mechanical properties. The RF shows greatest predictive power than other algorithms on the fracture strength (R = 0.9725, RRMSE = 23.56%), whereas the ANN algorithm gives the best results for the fatigue strength (R = 0.9699, RRMSE = 24.49%), the tensile strength (R = 0.9857, RRMSE = 16.89%), and hardness (R = 0.9836, RRMSE = 18.13%).

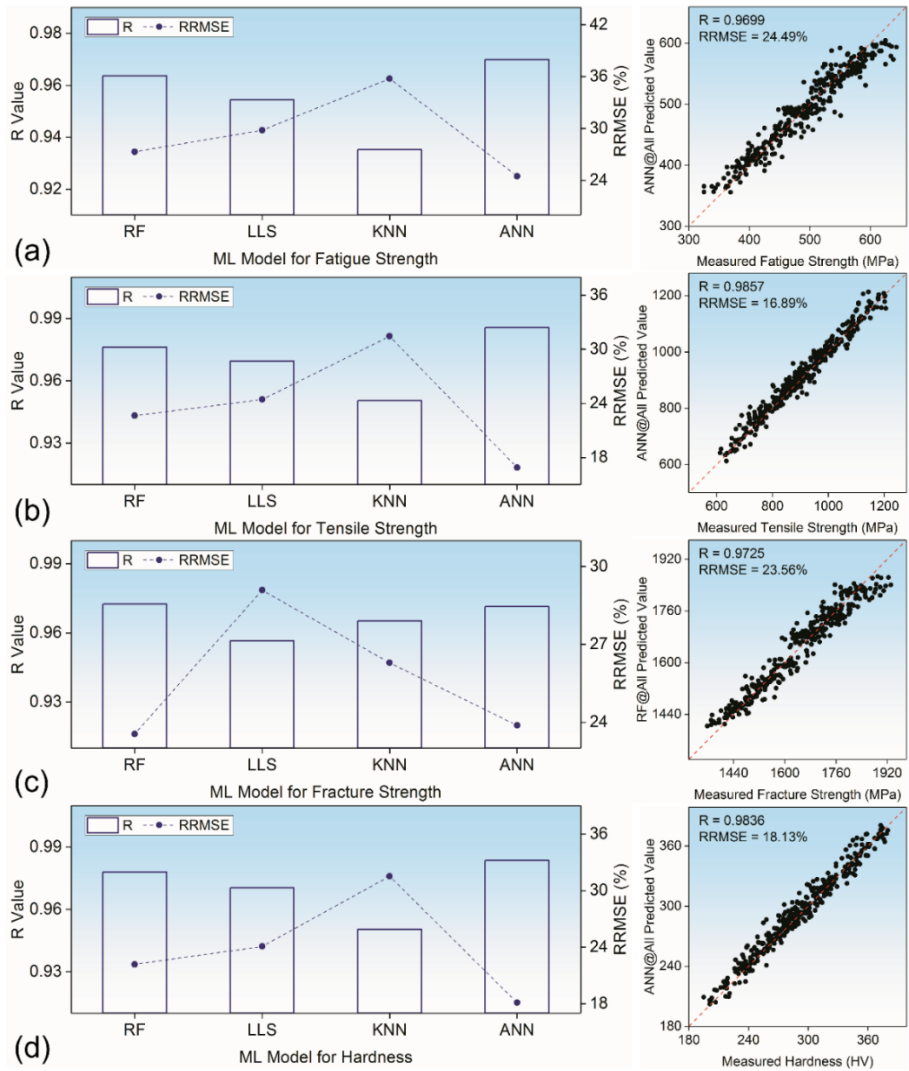

Figure 1. With all the 16 features, the R and RRMSE values of the RF, LLS, KNN, and ANN models for (a) fatigue strength and the performance of the best model ANN@All; (b) for tensile strength and the performance of the best model ANN@All; (c) for fracture strength and the performance of the best model RF@All; and (d) for hardness and the performance of the best model ANN@All.

## 3.2 Feature Selection

Feature selection is crucial in ML. ML algorithms, such as RF and symbolic regression (SR), have the feature selection functions and thus are emphasized here. The importance of features computed by RF was termed as RFI, and computed by SR was denoted by SRI. Figures 2 (a)-(b) show the RFI and SRI values of each original feature, respectively. The RFI values indicate that the top four important features are molybdenum, chromium, normalizing temperature, and tempering temperature, whereas the SRI values show the top four important features of tempering temperature, carbon, chromium, and molybdenum, which correspondingly yield two feature subsets of RFI (NT, TT, Cr, Mo) and SRI (TT, C, Cr, Mo).

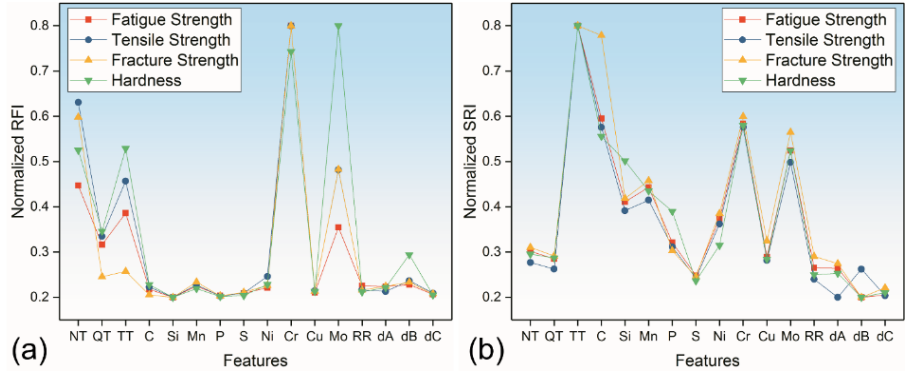

Figure 2. The normalized (a) random forest importance (RFI) and (b) symbolic regression importance (SRI) of the 16 features for fatigue strength, tensile strength, fracture strength, and hardness.

The four ML algorithms are conducted with the RFI (NT, TT, Cr, Mo) and SRI (TT, C, Cr, Mo) features. Figure 3 shows the cross-validation R-values and the predicted values of the best model against the measured value for each of the four properties. The results illustrate that the RF algorithm with the feature subset SRI (TT, C, Cr, Mo) outperforms other algorithms. The RF models with the feature subset SRI (TT, C, Cr, Mo) predict the four target properties with high predictive accuracy (R > 0.9550, RRMSE < 30.00%).

### 3.3 Mathematical Expressions

With SRI (TT, C, Cr, Mo) features, SR gave the following mathematical expressions for Fatigue Strength (FaS) in MPa, Tensile Strength (TS) in MPa, Fracture Strength (FrS) in MPa, and Hardness (H) in HV.

$$FaS = -0.8685TT + 316.7C + 367.6Cr - 227.5Cr^2 + 708.6Mo^2 + 785.0 \tag{3}$$

$$TS = -1.827TT - 119.7/C + 643.2Cr - 379.9Cr^2 + 1514Mo^2 + 2122 \tag{4}$$

$$FrS = -1.176TT - 46.12/C + 695.4Cr - 415.3Cr^2 + 1461Mo^2 + 2267 \tag{5}$$

$$H = -0.5839TT - 38.41/C + 191.2Cr - 113.3Cr^2 + 104.0Mo + 681.9 \qquad (6)$$

where all elements are in wt.% and TT is in (°C). Those equations show strong predictive power (R > 0.9425, RRMSE < 33.30%), as shown in Figure 4. In each of Eqs. (3-6), there is a minus sign with the tempering temperature, meaning that lower tempering temperatures are suggested to improve the strengths and hardness of steels. The alloying elements of carbon, chromium and molybdenum are also strengthening elements.

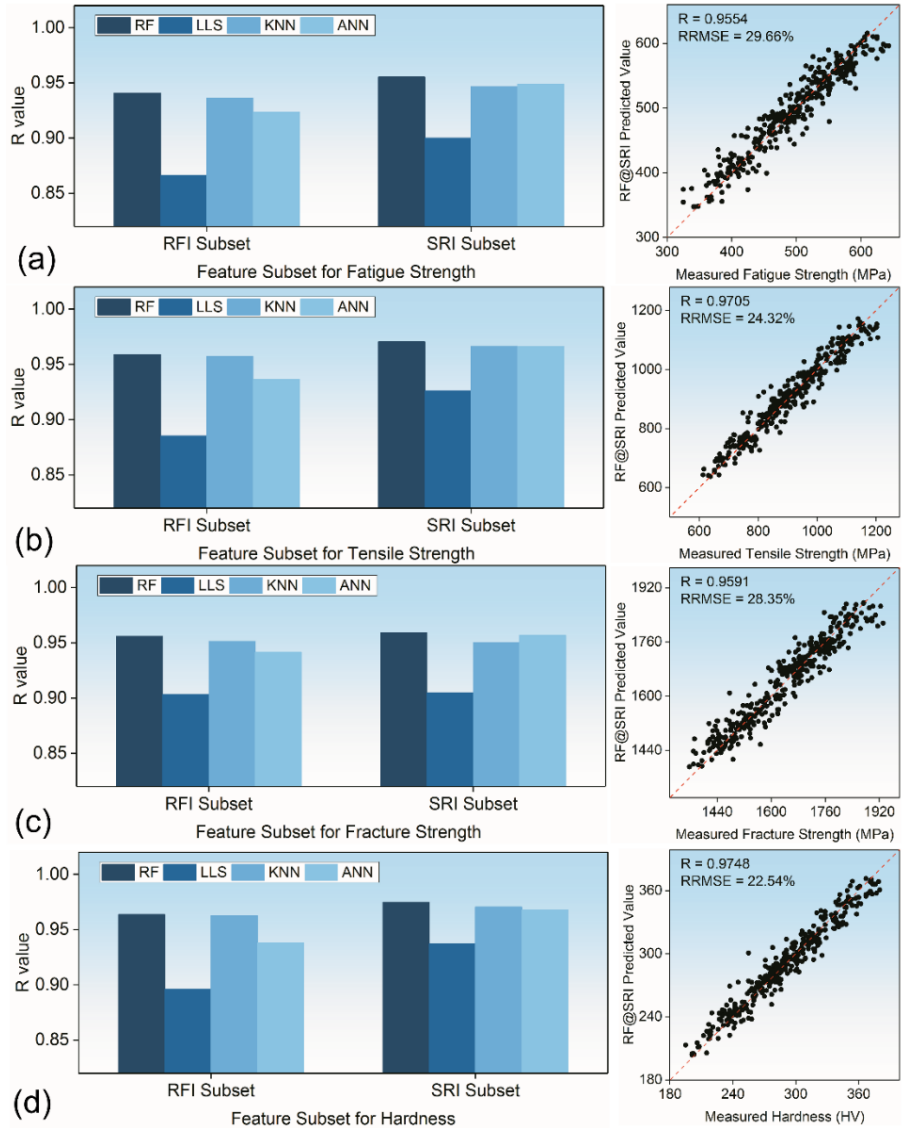

Figure 3. The R values of the RF, LLS, KNN, and ANN models with the selected RFI and SRI feature subsets (a) for fatigue strength and the performance of the best model RF@SRI; (b) for tensile strength and the performance of the best model RF@SRI; (c) for fracture strength and the performance of the best model RF@SRI; and (d) for hardness and the performance of the best model RF@SRI.

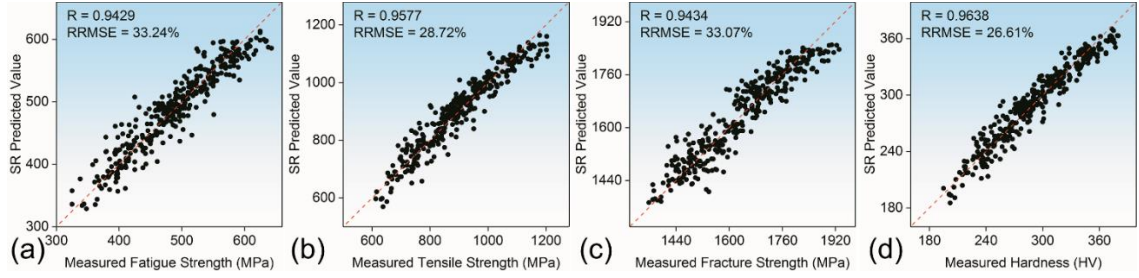

Figure 4. The performance illustrations of (a) Eq. (3) for fatigue strength, (b) Eq. (4) for tensile strength, (c) Eq. (5) for fracture strength, and (d) Eq. (6) for hardness.

### 3.4 ML model based on atomic features

To generalize the predication power of ML for new alloy discovery, atomic features are introduced in the present work. Iron is the matrix of steels, alloying elements in steels may behave as solutes, forming metal carbides with carbon, forming intermetallic compounds with iron or/and among alloying elements themselves, precipitation as clusters or/and tiny sized phases, etc. Table 2 lists the atomic features. All these atomic features and tempering temperatures are denoted by All-AF and used in the following ML.

In Table 2, $r_i$ and $r_{Fe}$ denote the atomic radii of element i and iron, respectively; $VEC_i$, $VEC_{Fe}$, and $VEC_C$ are the valance electrons of element i, iron, and carbon, respectively; $\chi_i$, $\chi_{Fe}$, and $\chi_C$ are the Pauling electronegativities of element i, iron, and carbon, respectively. Table S1 in the Supplementary Material list the values of these atomic properties. Besides, $a_i$ is the atomic percentage of element i, which links to the weight percentage $x_i$ by $a_i = \frac{x_i/M_i}{\sum_i(x_i/M_i)}$, where $M_i$ is the atomic weight of element i.

The feature selection of atomic features is also conducted by RF and SR. The RFI selects the three important atomic features of $t_{VEC}$, $d_{VEC\text{-}Fe}$, $d_{VEC\text{-}C}$, and TT for fatigue strength and hardness, and $t_{VEC}$, $a_{Fe}$, $d_{VEC\text{-}C}$, and TT for tensile strength and fracture strength. The SRI selects the four features of $d_{VEC\text{-}C}$, $d_{r\text{-}Fe}$, $a_{Fe}$, and TT for all the four mechanical properties. The selected features by RF and SR are referred as RFI-AF (TT, $t_{VEC}$, $d_{VEC\text{-}Fe}$, $d_{VEC\text{-}C}$), RFI-AF (TT, $t_{VEC}$, $a_{Fe}$, $d_{VEC\text{-}C}$), and SRI-AF (TT, $a_{Fe}$, $d_{r\text{-}Fe}$, $d_{VEC\text{-}C}$), respectively.

The RF algorithm is conducted again with All-AF, RFI-AF and SRI-AF features. Figure 5(a) shows the R-values for each of the four properties. The results indicate that the RF model with SRI-AF (TT, $a_{Fe}$, $d_{r\text{-}Fe}$, $d_{VEC\text{-}C}$) performs similar to the RF model with All-AF, and the RF model with SRI-AF performs super than the RF model with the two RFI-AF feature sets. Figures 5(b)¬(e) show the predicted values, by the RF model with SRI-AF, against the measured values for the four mechanical properties, respectively, and all R > 0.9510 and all RRMSE < 31.00%.

Table 2. Atomic features used in this work

| Features | Description | Formula |
|---|---|---|
| $a_{Fe}$ | Atomic percentage of Iron | $a_{10}$ |
| $t_r$ | Total atomic radius | $\sum_{i=1}^{10} a_i r_i$ |
| $d_{r\text{-}Fe}$ | Atomic radius difference (Iron-based) | $\sqrt{\sum_{i=1}^{10} a_i \left(1 - r_i/r_{Fe}\right)^2}$ |
| $t_{VEC}$ | Total Valance Electron | $\sum_{i=1}^{10} a_i VEC_i$ |
| $d_{VEC\text{-}Fe}$ | Valance Electron difference (Iron-based) | $\sqrt{\sum_{i=1}^{10} a_i \left(1 - VEC_i/VEC_{Fe}\right)^2}$ |
| $d_{VEC\text{-}C}$ | Valance Electron difference (Carbon-based) | $\sqrt{\sum_{i=1}^{10} a_i \left(1 - VEC_i/VEC_C\right)^2}$ |
| $t_\chi$ | Total Pauling Electronegativity | $\sum_{i=1}^{10} a_i \chi_i$ |
| $d_{\chi\text{-}Fe}$ | Electronegativity difference (Iron-based) | $\sqrt{\sum_{i=1}^{10} a_i \left(1 - \chi_i/\chi_{Fe}\right)^2}$ |
| $d_{\chi\text{-}C}$ | Electronegativity difference (Carbon-based) | $\sqrt{\sum_{i=1}^{10} a_i \left(1 - \chi_i/\chi_C\right)^2}$ |

Similarity, Equations (7-10) from SR give the explicit correlations of Fatigue Strength (FaS) in MPa, Tensile Strength (TS) in MPa, Fracture Strength (FrS) in MPa, and Hardness (H) in HV, with the SRI-AF, respectively.

$$FaS = -0.8631TT - 2771a_{Fe} + 6679d_{r-\text{Fe}} + 27690d_{\text{VEC}-\text{C}} - 10610 \quad (7)$$

$$TS = -1.801TT - 4438a_{Fe} + 14852d_{r-\text{Fe}} + 58552d_{\text{VEC}-\text{C}} - 24019 \quad (8)$$

$$FrS = -1.148TT - 4718a_{Fe} + 9863d_{r-\text{Fe}} + 60564d_{\text{VEC}-\text{C}} - 24003 \quad (9)$$

$$H = -0.5724TT - 1122a_{Fe} + 4810d_{r-\text{Fe}} + 18906d_{\text{VEC}-\text{C}} - 8062 \quad (10)$$

where $a_{Fe}$ is in at.% and TT is in (°C). Those equations indicate that the alloying elements enhances the strengths of steels. Figures 6 (a-d) show the performances of Equations (7-10), respectively, and associated with the R and RRMSE values.

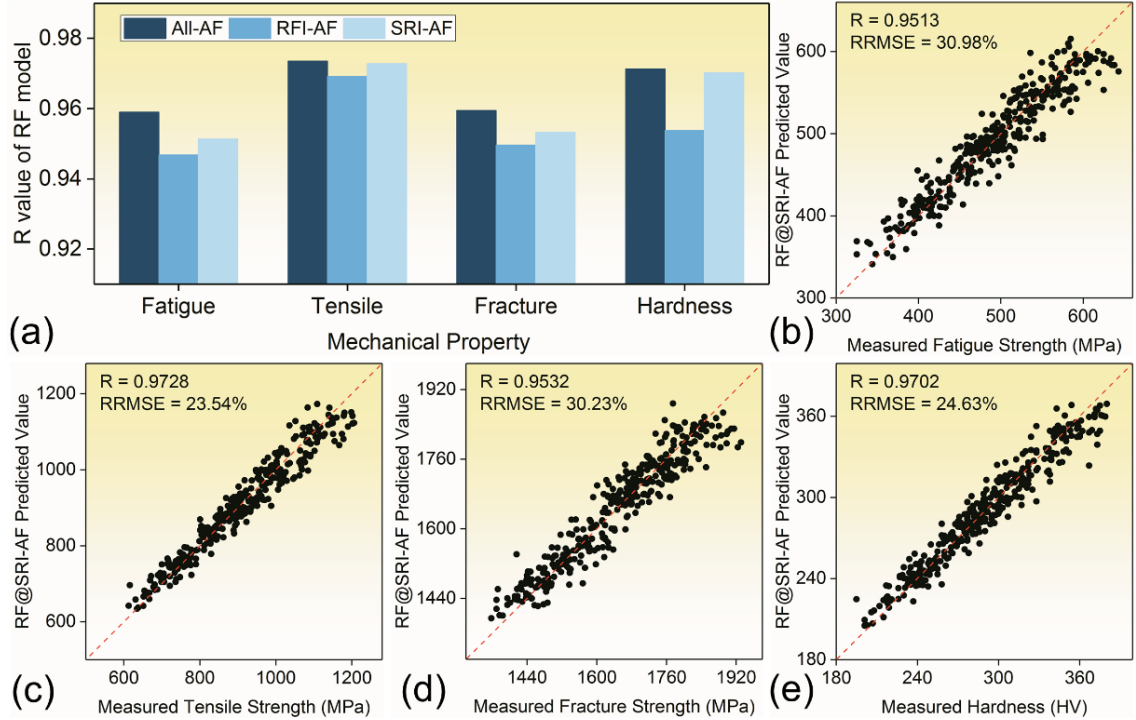

Figure 5. (a) The R-values of the RF models with All-AF, RFI-AF, and SRI-AF features for the four mechanical properties. The predicted values of RF model with SRI-AF features against the measured values for (b) fatigue strength, (c) tensile strength, (d) fracture strength, and (e) hardness.

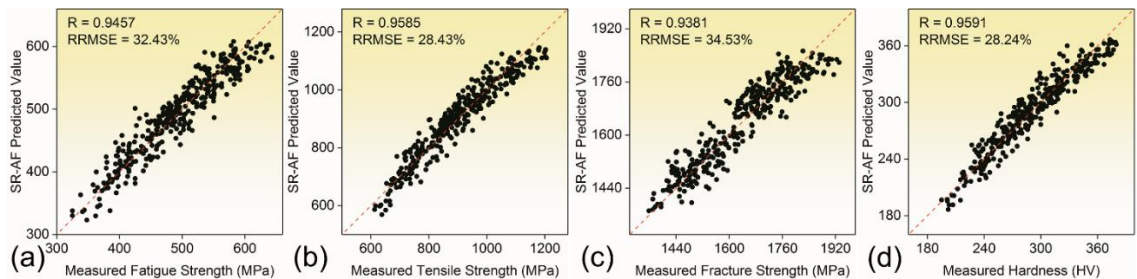

Figure 6. The performance illustrations of (a) Eq. (7) for fatigue strength, (b) Eq. (8) for tensile strength, (c) Eq. (9) for fracture strength, and (d) Eq. (10) for hardness.

## 3.5 Development of anti-fatigue high strength steel

In the used 360 data, the lowest tempering temperature is 550 ºC for forming tempering sorbate and the maximum contents of C, Cr, and Mo are 0.57 wt. %, 1.12 wt. %, and 0.24 wt. %, respectively. Thus, a novel anti-fatigue high strength steel is possibly discovered with the heat treatment condition and compositions shown in Table 3 with the ML predicted mechanical properties shown in Table 4. Although the ML predictions from Eqs. (3-6) deviate slightly from these corresponding values from Eqs. (7-10), the average predicted fatigue strength $682.5 \pm 27.5$ MPa at fatigue life of $10^7$, tensile strength of $1286 \pm 48$ MPa), and hardness $406 \pm 16$ HV are all higher than the corresponding maximum values, and the average predicted fracture strength $1922 \pm 41$ MPa is comparable to the maximum fracture strength reported 1931 MPa.

Table 3. The tempering temperature and composition of the data-driven discovered anti-fatigue high strength steel

| TT | C | Cr | Mo | Other Features |
|---|---|---|---|---|
| 550 ºC | 0.57 wt% | 1.12 wt% | 0.24 wt% | Maximum value (minimize $a_{Fe}$) |

Table 4. The four mechanical properties of the data-driven discovered anti-fatigue high strength steel

| Properties | Maximum value in the dataset | Predictions of Eq. (3)-(6) | Predictions of Eq. (7)-(10) | Average Predictions |
|---|---|---|---|---|
| FaS (MPa) | 643 | 655 | 710 | $682.5 \pm 27.5$ |
| TS (MPa) | 1206 | 1238 | 1334 | $1286 \pm 48$ |
| FrS (MPa) | 1931 | 1881 | 1963 | $1922 \pm 41$ |
| H (HV) | 380 | 390 | 422 | $406 \pm 16$ |

## 4. Concluding Remarks

ML and feature selection are conducted on 360 data of steels to predict the fatigue strength at fatigue life of $10^7$ cycles, tensile strength, fracture strength, and hardness of steels; and to find the most important features to the four mechanical properties. The ML results demonstrate that the tempering temperature and the elements of carbon, chromium, and molybdenum are the key features to the mechanical properties of steels, with which the RF model exhibits the high validation accuracy, viz. R > 0.9550, RRMSE < 30.00%. In particular, the SR gives explicitly mathematic expressions of the four mechanical properties as functions of the four important features, and hence accordingly, designs an anti-fatigue high strength steel.

**Method and Software**

Four ML algorithms (RF, LLS, KNN, ANN) in the WEKA software library [17] and symbolic regression algorithm in the HeuristicLab [18] were used in the present work. In the open source software, all parameters of ML algorithms were set as the default, unless otherwise requested.

**RF**: The number of features randomly chosen at each node is denoted by *numFeatures* and determined via grid search to achieve the highest predicting accuracy. The search results are shown in Table 5 for each feature subset. The RFI value was computed based on the mean decrease impurity [19] in WEKA.

Table 5. The number of features randomly chosen for each subset

| Training set | All | RFI | SRI | All-AF | RFI-AF | SRI-AF |
| --- | --- | --- | --- | --- | --- | --- |
| *numFeatures* | 7 | 1 | 2 | 5 | 2 | 2 |

**KNN**: The number of neighbours is denoted by *KNN* and determined via grid search, *KNN* is recommend to be 4, 2, and 3 for All, RFI, and SRI feature subsets, respectively.

**ANN**: The number of hidden layers in the neural network and the learning rate of weight update are denoted by *hiddenLayers* and *learningRate*, respectively. The two hyper-parameters were determined via grid search to be *learningRate* = 0.1 and the *hiddenLayers* of 8, 7, and 7 for All, RFI, and SRI feature subsets, respectively.

**SR**: Genetic Programming (GP) in Heuristic Lab was utilized to search for an optimal expression. The parameters of GP used in the present work are listed in Table 6. One hundred independent GP runs are conducted in this work. The SRI value was computed as the fitness-weighted variable importance described as defined in [20] based on 100 individual GP runs.

Table 6 The used parameters in GP

| Parameter | *Population Size* | *Number of Generation* | *Mutation Probability* | *Crossover Probability* | *Maximum Tree Depth* | *Maximum Tree Length* |
|---|---|---|---|---|---|---|
| Value | 1000 | 10000 | 20% | 80% | 10 | 15 |

**Acknowledgments**

The work is supported by the National Key R&D Program of China (No. 2018YFB0704404), the Hong Kong Polytechnic University (internal grant nos. 1-ZE8R and G-YBDH), and the 111 Project of the State Administration of Foreign Experts Affairs and the Ministry of Education, China (grant no. D16002).

**References**

1 Ramprasad R, Batra R, Pilania G, et al. Machine learning in materials informatics: Recent applications and prospects. npj Comput Mater, 2017, 3: 54.

2 Xue D, Xue D, Yuan R, et al. An informatics approach to transformation temperatures of NiTi-based shape memory alloys. Acta Mater, 2017, 125: 532–541.

3 Ward L, Agrawal A, Choudhary A, et al. A general-purpose machine learning framework for predicting properties of inorganic materials. npj Comput Mater, 2016, 2: 1–7.

4 Senderowitz H, Barad H-N, Yosipof A, et al. Data Mining and Machine Learning Tools for Combinatorial Material Science of All-Oxide Photovoltaic Cells. Mol Inform, 2015, 34: 367–379.

5 Agrawal A, Choudhary AN. Perspective: Materials informatics and big data: Realization of the "fourth paradigm" of science in materials science. APL Mater, 2016, 4: 53208.

6 Xiong J, Shi SQ, Zhang TY. A machine-learning approach to predicting and understanding the properties of amorphous metallic alloys. Mater Des, 2020, 187: 108378

7 Takahashi K, Tanaka Y. Material synthesis and design from first principle calculations and machine learning. Comput Mater Sci, 2016, 112: 364–367.

8 Seshadri R, Wolverton C, Hill J, et al. Materials science with large-scale data and informatics: Unlocking new opportunities. MRS Bull, 2016, 41: 399–409.

9 Green ML, Choi CL, Hattrick-Simpers JR, et al. Fulfilling the promise of the materials genome initiative with high-throughput experimental methodologies. Appl Phys Rev, 2017, 4: 011105.

10 Huber L, Hadian R, Grabowski B, et al. A machine learning approach to model solute grain boundary segregation. npj Comput Mater, 2018, 4: 64.

11	Zhu Q, Samanta A, Li B, et al. Predicting phase behavior of grain boundaries with evolutionary search and machine learning. Nat Commun, 2018, 9: 467.

12	Falk C, Wenny MB, Norquist AJ, et al. Machine-learning-assisted materials discovery using failed experiments. Nature, 2016, 533: 73–76.

13	Agrawal A, Deshpande PD, Cecen A, et al. Exploration of data science techniques to predict fatigue strength of steel from composition and processing parameters. Integr Mater Manuf Innov, 2014, 3: 90-108.

14	Agrawal A, Choudhary A. An online tool for predicting fatigue strength of steel alloys based on ensemble data mining. Int J Fatigue, 2018, 113: 389-100

15	Yamazaki M, Xu Y, Murata M, et al. NIMS structural materials databases and cross search engine - MatNavi. VTT Symp, 2007.

16	Lison P. An introduction to machine learning. Language Technology Group: Edinburgh, UK, 2015.

17	Hall M, Frank E, Holmes G, et al. The WEKA data mining software: an update. ACM SIGKDD Explor Newsl, 2009, 11(1): 10-18.

18	Wagner S, Affenzeller M. HeuristicLab: a generic and extensible optimization environment. Adapt Nat Comput Algorithms. Vienna: Springer, 2005. 538-541.

19	Louppe G, Wehenkel L, Sutera A, et al. Understanding variable importances in forests of randomized trees. Adv Neural Inf Process Syst, 2013. 431-439.

20	Vladislavleva K, Veeramachaneni K, Burland M, et al. Knowledge mining with genetic programming methods for variable selection in flavor design. In: Proc 12th Annu Genet Evol Comput Conf, GECCO 2010. 941-948.